\documentclass[twocolumn,abs,prb,showpacs]{revtex4-1}

\usepackage{graphicx}
\usepackage{dcolumn}
\usepackage{float}
\usepackage{amsmath}
\usepackage{bm}

\newcommand{\comment}[1]{}

\begin{document}


\title{Comparison of pressurized sulfur hydride with conventional superconductors}

\author{E. J. Nicol$^{1}$}
\email{enicol@uoguelph.ca}
\author{J. P. Carbotte$^{2,3}$}
\affiliation{$^1$Department of Physics, University of Guelph,
Guelph, Ontario N1G 2W1, Canada} 
\affiliation{$^2$Department of Physics and Astronomy, McMaster
University, Hamilton, Ontario L8S 4M1, Canada}
\affiliation{$^3$The Canadian Institute for Advanced Research, Toronto, ON M5G 1Z8, Canada}
\date{\today}

\begin{abstract}{
A recent report that sulfur hydride under pressure is an electron-phonon 
superconductor with a $T_c$ of 190~K has been met with much excitement 
although it is yet to be confirmed. Based on several electron-phonon spectral 
density functions already available from density functional theory, we find that 
the electron-phonon spectrum is near optimum for $T_c$ with a particularly large value
of its characteristic phonon energy $\omega_{\rm ln}$ which is due to the small
hydrogen mass. We find that the thermodynamic universal BCS ratios are near
those for Pb and Nb$_3$Sn. We suggest that optical measurements could be a useful tool
to establish the existence and nature of the superconductivity in this system.
Conventional superconductors are in the impurity-dominated dirty limit. By 
contrast sulfur hydride will be in the clean limit because of its large energy gap
scale. The AC optical conductivity will display distinct and separate 
signatures of the superconducting gap in the low-energy impurity-dominated range
of the optical spectrum and additional phonon structures at higher energies 
where the clean limit applies. 
}
\end{abstract}



\maketitle


 The observation\cite{Drozdov:2014} of a dramatic and sharp drop in resistivity 
around a temperature of $T\lesssim 190 K$ in a sulfur hydride sample under
$\sim 200$ GPa, has led to a flurry of new theoretical works\cite{Duan:2014,Errea:2015,Flores-Livas:2015,Mazin:2015,Pickett:2015,Hirsch:2015,Akashi:2015}. The 
verification that it is definitely superconducting, for example through the 
observation  of a Meissner effect, still awaits further experiments, however,
the authors
did observe that with the application of a magnetic field of 7 Tesla there was a
shift,  to lower
temperature, of the  onset of the resistivity drop by $\sim 10$ K. Moreover, the
conductivity below 190 K was found to be better than that of the purest metal
at low temperature and was two orders of magnitude larger than that of pure copper.

A number of papers\cite{Duan:2014,Errea:2015,Flores-Livas:2015} 
based on density functional theory for sulfur hydride 
have already provided detailed information on the electron-phonon spectral
density function $\alpha^2F(\omega)$ and  have identified H$_3$S as the 
most likely phase involved at that pressure. While some quantitative 
differences exist between the
various spectral densities available, all calculations  support the
idea of phonon-mediated superconductivity in H$_3$S with a $T_c$ of order 200~K
at $\sim 200$ GPa. Still it remains essential to find ways to directly verify
such a possibility. Lacking definitive confirmation, it is instructive
to place H$_3$S within the context of the family of known conventional 
superconductors\cite{Carbotte:1990}. In particular an emphasis on differences 
and commonalities between H$_3$S and the well-established cases of 
electron-phonon-mediated superconductivity can provide valuable insight. It is also
important to calculate the expected properties of the superconducting state of
H$_3$S with the specific aim of providing information that can be
used by experimentalists to establish that H$_3$S is indeed a superconductor
and to discover the pairing mechanism involved, {\it i.e.}, the glue that
binds the Cooper pairs.

In this Rapid Communications, we demonstrate that H$_3$S is a very highly optimized
electron-phonon superconductor for various reasons and we provide a calculation
of the optical conductivity as a test for identifying the superconducting
gap and the Holstein structure for the phonon spectrum. For this purpose, we
work with the $\alpha^2F(\omega)$ spectrum provided in the literature. For illustrative
purposes we have used the one developed by Errea {\it et al.}\cite{Errea:2015}
(shown in Fig.~\ref{fig1}) which
includes anharmonic phonon effects. 
However, in the Supplemental Material\cite{supp} (Fig.~S1), we also show the spectra for the harmonic version
from Errea {\it et al.} 
along with that from Flores-Livas {\it et al.}\cite{Flores-Livas:2015} 
for comparison [the one from Duan {\it et al.}\cite{Duan:2014} is 
also similar but not shown]. 
Using the $\alpha^2F(\omega)$, one can calculate
the electron-phonon mass renormalization parameter, 
$\lambda=\int_0^\infty[\alpha^2F(\omega)/\omega]d\omega$, the area under the
spectrum $A=\int_0^\infty\alpha^2F(\omega)d\omega$ and the Allen-Dynes
characteristic phonon frequency\cite{Allen:1975,McMillan:1968} $\omega_{\rm ln}={\rm exp}[(2/\lambda)\int_0^\infty[\ln(\omega)\alpha^2F(\omega)/\omega]d\omega]$, which would equal $\omega_E$ if the
$\alpha^2F(\omega)=A\delta(\omega-\omega_E)$. For our digitization of the published spectrum shown in Fig.~\ref{fig1},
$\lambda=1.67$, $A=118.5$ meV, and $\omega_{\rm ln}=122$ meV.
Comparison of these characteristic properties with the other
spectra are given in Table~S1 of the 
Supplemental Material\cite{supp}. 
[Note, these
spectra were digitized from the original papers and placed on a
2 meV grid. This gives characteristics that are in qualitative agreement
with the original works if not always in perfect quantitative agreement.
Such details do not impact our results and conclusions presented here.
It is clear that all of the spectra are qualitatively similar.\cite{supp}]
Calculating $T_c$ with these
spectra using the standard s-wave electron-phonon Eliashberg equations\cite{Carbotte:1990}, 
we find that for $T_c=190$ K, the Coulomb pseudopotential
$\mu^*$ would have to be 0.18 for
the anharmonic spectrum and about 0.38-0.4 for the harmonic spectrum (or conversely, 
for a $\mu^*\sim 0.16$, the harmonic spectra would give a $T_c$ of order
250 K, which has been noted by the authors of these spectra). While a large
$\mu^*$ is unusual, such a value has been found for the alkali-doped fullerenes\cite{Marsiglio:1998} and can
be taken as an argument for strong electron correlations to be important in addition to 
the electron-phonon interaction. It should be noted that the spectrum of 
Flores-Livas
is one that has been developed and used by accounting for a nonconstant energy-dependent electronic
density of states and yet it essentially agrees with the Errea harmonic spectrum. These details aside and working with the anharmonic spectrum
for illustration,
we argue that the $\alpha^2F(\omega)$ for H$_3$S is one of the most highly optimized among
conventional superconductors. 

One test of this concept is to evaluate the functional derivative of $T_c$ [Ref.~\cite{Mitrovic:1981}] with
respect to this spectrum as is shown in Fig.~\ref{fig1}(a) (red curve). This
derivative is peaked at $\omega\sim 7k_BT_c$ and illustrates that if one
could move all the spectral weight in the $\alpha^2F(\omega)$ to a delta function placed at this frequency, then one would obtain the highest $T_c$ possible from
this spectrum. However, the spectrum is distributed around this point and so the question
remains as to how optimized is it? Using the characteristic phonon energy $\omega_{\rm ln}$
as a measure, one finds that it is very close to the value of $7T_c$ as shown
in the figure. Consequently, the spectrum is quite highly optimized already. This is
understood from the fact that the functional derivative varies quite 
slowly about its peak and much of the weight of the $\alpha^2 F(\omega)$ falls in the
region of this slow variation. The shape of the functional derivative speaks
to the physics that for phonon frequencies that are too low, the phonons do
not respond fast enough to pair the electrons, but rather act more like
static impurities. Alternatively, at very high frequencies, the ions would wiggle too
fast to have any net displacement or polarization to glue the Cooper pairs.
As a result the optimal frequency is found at some intermediate frequency.
A simple physical argument for the maximum to occur at $\omega\sim 7k_BT_c$ can be found in Refs.~\cite{Carbotte:1987,Carbotte:1990}

The question as to how optimal this spectrum is can be quantified by
examining Fig.~\ref{fig1}(b) where the dimensionless ratio of $T_c/A$ is plotted
versus $\mu^*$. The line with the open circles is the maximum value that $T_c$ could take for fixed $A$ as a function of $\mu^*$ and is calculated by placing a delta function
of weight $A$ at the frequency of the peak in its own functional derivative and finding the 
$T_c$. This is a classic curve taken from Leavens\cite{Leavens:1975} and the solid
black dots represent the data for several conventional superconductors taken from
a tabulation (Table IV) in Ref.~\cite{Carbotte:1990}. The gold star shows where the spectrum
in Fig.~\ref{fig1}(a) sits and it is just below the optimal line. Indeed, it
falls only 8\% below the theoretical absolute upper bound allowed to any spectrum with that area $A$ and $\mu^*$.
 Furthermore,
the blue curve, shows how $T_c$ would vary for this spectrum as a function of
$\mu^*$ and it is clear that it would always remain close to the upper bound. For $\mu^*=0$, the $T_c= 298$ K.

 \begin{figure}
\includegraphics[width=0.9\linewidth]{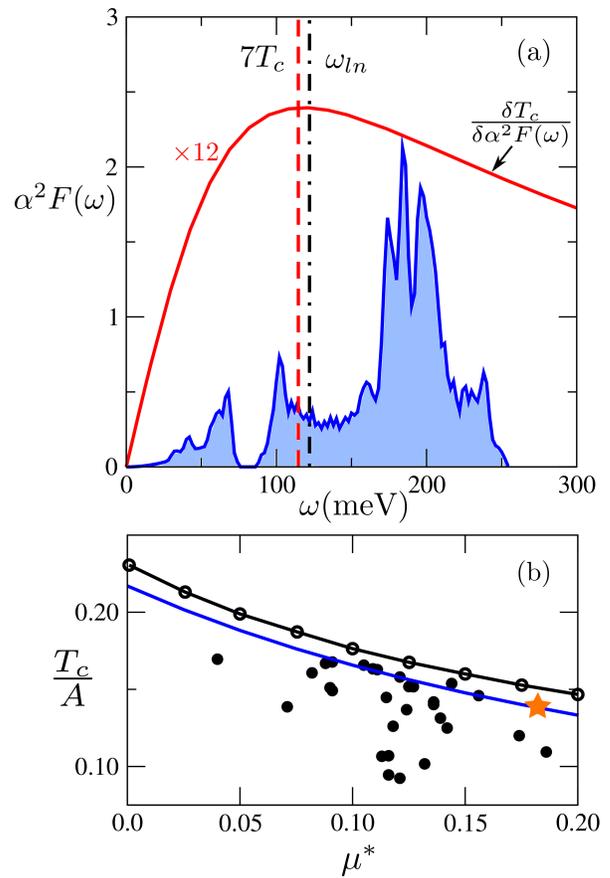}
\caption{(Color online) 
(a) The anharmonic $\alpha^2F(\omega)$ spectrum for H$_3$S at 200 GPa given
in Ref.~\cite{Errea:2015}. The functional derivative 
$\delta T_c/\delta\alpha^2 F(\omega)$ vs $\omega$ 
for this same spectrum is also 
shown scaled up by a factor of 12. It peaks at $\omega=7k_BT_c$ 
(dashed line) which is very close to the characteristic phonon frequency 
$\omega_{\rm ln}$ (dash-dotted line) for the $\alpha^2F(\omega)$ 
demonstrating that the spectrum is highly optimized for $T_c$.
(b) $T_c/A$ vs $\mu^*$ where $A$ is the area under the $\alpha^2F(\omega)$ 
spectrum. The black curve with open circles is the theoretical upper limit for
$T_c$ with fixed $A$  
and the solid circles are data for several superconductors constructed from Table~IV
of Ref.~\cite{Carbotte:1990}. The gold star is for the $\alpha^2F(\omega)$
spectrum in (a) with $T_c=190$ K and $\mu^*=0.18$. The blue curve represents how
$T_c/A$ would change for varying $\mu^*$, demonstrating that this spectrum is near optimal.
}\label{fig1}
\end{figure}

One may now try to understand from this why the $T_c$ is so high.
 It is the unusually large value of 
$A$ in H$_3$S which accounts for the large $T_c$ found in this system. This sizable
$A$ comes about because the electron-ion coupling remains large to very high 
phonon energies. In turn, the unusually large value of $\omega_{\rm ln}$ is
traced to the small hydrogen mass and the stiffness of the H-S bond stretching
mode. Moreover, the lack of an electronic core structure to the hydrogen 
(see particularly Ref.~\cite{Flores-Livas:2015}) means
that there is no orthogonalization of the conduction electron wavefunctions which in
ions with core electrons effectively changes the electron-ion interaction
potential into a weakened pseudopotential. These two facts combined provide
the mechanism for a large area $A$ under the spectral density $\alpha^2F(\omega)$
which extends to high energies with large magnitude. Further insight into the
importance of a large $\omega_{\rm ln}$ can be obtained through an analysis
based on the asymptotic formula\cite{Carbotte:1990} for $T_c$ of Eliashberg theory derived on the
assumption of a very large value of $\lambda$, the mass enhancement factor. In
this limit $T_c=0.182\sqrt{\lambda}\omega_{\rm ln}$. We see that increasing $\lambda$
by a factor of 2 increases $T_c$ by $\sim 40$\% while increasing $\omega_{\rm ln}$ by 2 doubles the $T_c$. While H$_3$S is not in the asymptotic limit of large
$\lambda$, this analysis nevertheless provides a useful qualitative picture.
While $\lambda$ for H$_3$S is not particularly large as compared with other
electron-phonon superconductors, $\omega_{\rm ln}$ is quite significant at
about 120 meV. Through the numerical solution of the Eliashberg equations
for many model spectra, Leavens and Carbotte\cite{Leavens:1974} observed that,
for $\lambda$ and $\mu^*$ in the range of $1.2<\lambda<2.4$ and $0.1<\mu^*<0.15$, a 
remarkably simple relationship between $T_c$ and $A$ applied, 
namely $T_c=0.148 A$. This formula 
can be used to obtain further understanding of the role various regions of 
the $\alpha^2F(\omega)$ spectrum play in $T_c$. As an example, the entire 
region below $\sim 165$ meV in the spectrum of Fig.~\ref{fig1}(a) has an area of 39 meV while the area above is 80 meV,
contributing, respectively, 67 and 137 K to the $T_c$ value. It is clear that
the hydrogen bending vibrations and most importantly the stretching modes are 
critical for the high $T_c$ of H$_3$S metallic phase.

\begin{figure}
\includegraphics[width=0.9\linewidth]{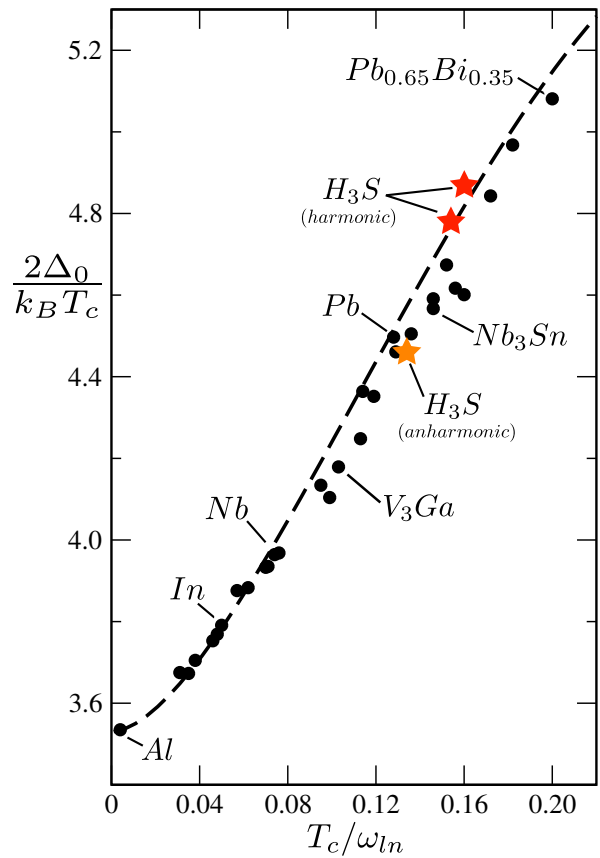}
\caption{(Color online) The gap ratio $2\Delta_0/k_BT_c$ vs $T_c/\omega_{\rm ln}$. The solid black points are the data for several conventional superconductors
 taken from Table IV of Ref.~\cite{Carbotte:1990}. Some have been labeled. The dashed line is an analytical formula developed for conventional superconductors
from strong coupling Eliashberg theory\cite{Mitrovic:1984}. The
gold star is the value calculated for the anharmonic spectrum shown in
Fig.~\ref{fig1} and  the red stars are the values for the harmonic spectra\cite{supp}
with $T_c=190$ K. 
}\label{fig2}
\end{figure}

In Fig.~\ref{fig2}, we compare the gap ratio calculated for sulfur hydride with other conventional
superconductors on a classic plot\cite{Mitrovic:1984,Geilikman:1966} 
for the gap ratio versus $T_c/\omega_{\rm ln}$. We have done this calculation for both the anharmonic spectrum
(gold star) and the two harmonic spectra shown in the Supplemental Material\cite{supp} (with $T_c=190$ K).
Clearly, sulfur hydride fits well among the conventional superconductors and
sits close to Pb and Nb$_3$Sn which are other highly optimized superconductors.
Using this figure one can also translate to other figures for strong coupling
corrections to BCS ratios as found in Refs.~\cite{Marsiglio:1986,Carbotte:1990}. We have
done the numerical calculation and find 
for the anharmonic spectrum the specific heat ratio $\Delta C(T_c)/\gamma T_c =2.53$ and the magnetic field ratio $\gamma T_c^2/H_c^2(0)=0.136$, which are very close to the values found for Pb, as expected from Fig.~\ref{fig2}. 

We conclude from this that sulfur hydride appears to be a very highly optimized
electron-phonon superconductor which uses the high phonon 
frequency of the hydrogen and large electron-ion interaction to obtain a high $T_c$. Nonetheless, it remains a question as to how to demonstrate the superconductivity
and verify the electron-phonon interaction. Given that the sample is under pressure, many typical experimental probes may not be used. Tunneling was the
classic experiment in conventional superconductors but here the energy scales
are much higher than those that are usually accessed with tunneling. Consequently,
an experiment which is non-contact but can probe high energy scales is needed. Here one is further helped by the fact that the material will be in the ``clean limit'', where $2\Delta_0 \gg 1/\tau$ with $1/\tau$ being the impurity scattering
rate.\cite{Akis:1991} 
An excellent probe in this case, then, is optical conductivity measurements as they may be
done on a sample in a high pressure cell. As the cuprate superconductors have
already illustrated, the clean limit allows one to see boson structure in
optics.\cite{Carbotte:1999} 
This is illustrated in Fig.~\ref{fig3} where we show our calculation for the
optical conductivity for the anharmonic spectrum of H$_3$S. [A comparison with
that from the other $\alpha^2F(\omega)$ spectra is given\cite{supp} in Fig.~S2]. Shown
are the superconducting and normal state curves for $T=0.1T_c$. In the latter case, $\Delta$ is set to zero in the calculation\cite{Akis:1991}
 and this curve is shown at the same temperature
to illustrate differences between the two states independent of temperature. [For the normal
state at $T_c$, which is not that different, see Fig.~S3 of the Supplemental
Material\cite{supp}.] 
While the
normal state shows a Drude absorption followed by a bump associated with
the main source of inelastic scattering from the 200 meV peak in the electron-phonon
spectrum (see Fig.~\ref{fig1}), the superconducting state is much more dramatic.
Here one sees a sharp impurity-dominated peak at the gap edge of $2\Delta_0=73$
meV and then a series of absorption peaks. This structured absorption can be 
traced to the $\alpha^2F(\omega)$ spectrum which we show in Fig.~\ref{fig3}
shifted along the frequency axis by $2\Delta_0$.
A full analysis and inversion\cite{Carbotte:2011} is needed to pull out the $\alpha^2F(\omega)$
from the data, but it is clear that there are signatures of the energy gap, phonon structures and a shift in the main absorption hump going from the normal state 
to the superconducting state. By referring\cite{supp} to Fig.~S3 for temperature-dependent
spectra, it can be seen that such structures and overall response survive up to very
high temperature and are quite robust. As a final note, for these calculations
we have used a transport impurity scattering rate of 20 meV which is reduced by
a $1+\lambda$ factor\cite{Mori:2008} to make the halfwidth on the Drude conductivity 
to be about 7.5 meV in this
case. It should be remarked that the behavior seen here due to the clean limit
is quite different from what is found for conventional superconductors, such as
Pb,\cite{Mori:2008} which are in the dirty limit. The high energy scale of the superconducting gap contributes to sulfur hydride being in the clean limit and showing strong
Holstein structure.

\begin{figure}
\includegraphics[width=0.9\linewidth]{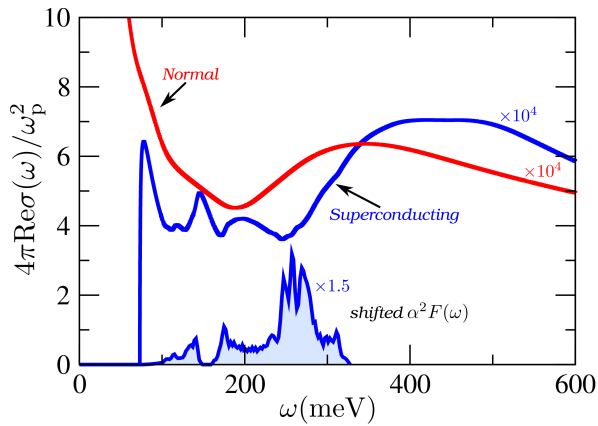}
\caption{(Color online)
The real part of the optical conductivity (scaled up by a factor of $10^4$)
vs $\omega$ at temperature $T=0.1T_c$. Shown are the normal and superconducting state
curves (red and blue, respectively). $\omega_p$ is the plasma frequency. The $\alpha^2F(\omega)$ spectrum from
Fig.~\ref{fig1} is shown shifted to the right along the frequency axis by $2\Delta_0$.  
}\label{fig3}
\end{figure}

In summary, using three different electron-phonon spectral densities now available 
in the literature which are based on density functional band structure computations
for H$_3$S under 200 GPa of pressure, we have verified that all give spectra that
are near optimum for $T_c$. More specifically all have a characteristic phonon
energy $\omega_{\rm ln}$ of Eliashberg theory which falls near the maximum
of its own functional derivative of $T_c$ with respect to $\alpha^2F(\omega)$. 
Given that the functional derivative $\delta T_c/\delta\alpha^2F(\omega)$ 
displays a very broad maximum, a significant distribution of phonon energies
about $\omega_{\rm ln}$ does not deplete $T_c$ much.
For example, in the case of the $\alpha^2F(\omega)$ provided by Errea {\it et al.}\cite{Errea:2015} including anharmonicity, the value of $T_c$ obtained falls only 8\% below its theoretical absolute maximum for any distribution of phonon
energies keeping the area under the spectral density fixed.

In Eliashberg theory, it was found that a critical parameter which measures 
coupling strength is the dimensionless ratio $T_c/\omega_{\rm ln}$.
For small values of this ratio BCS theory is recovered
while strong coupling corrections to the universal dimensionless ratio of BCS
increase with increasing value of $T_c/\omega_{\rm ln}$. For H$_3$S, one finds a value 
near 0.13 which is close to that found in Pb, Nb$_3$Sn and others. It corresponds to the
case where $\omega_{\rm ln}$ is close to the maximum in the functional derivative
$T_c/\omega_{\rm ln}\approx 7$ and hence accounts for the near optimization of the
spectrum. This also implies that the thermodynamic properties of H$_3$S will be close to those of Pb and Nb$_3$Sn. This does not imply, however, that other
properties of their  superconducting state will be similar. As an example, optical 
properties in H$_3$S will be very different from those in other conventional superconductors which are in the dirty (impurity-dominated) limit\cite{Mori:2008}. Instead, because of the
very large energy scale associated with its phonons and its superconducting gap value,
H$_3$S is in the clean (phonon-dominated) regime. To be more precise, the
absorptive part of the AC longitudinal conductivity ${\rm Re}\sigma_{xx}(\Omega)$
as a function of photon energy $\Omega$ will show two distinct and separate regimes.
At the energy of twice the gap, the regime remains impurity-dominated and 
there is a very sharp rise in  absorption at $\Omega=2\Delta$. but this is followed
by dominant phonon-assisted absorption  structures extending to many times $2\Delta$
which is usually referred to as the clean limit regime because residual impurity
scattering plays no significant role. This region provides a detailed picture
of the pairing glue. Consequently, optics can be used to provide
information not only on the existence of superconductivity but also of
the driving mechanism.


We thank T. Timusk for discussions, and B. Pavlovic and C. Tabert
for assistance with figures.
This work has been supported by the Natural Sciences and Engineering Research Council (NSERC) of Canada and, in part, by the
Canadian Institute for Advanced Research (CIFAR).


\bibliographystyle{apsrev4-1}
\bibliography{h3sbib}

\begin{thebibliography}{24}%
\makeatletter
\providecommand \@ifxundefined [1]{%
 \@ifx{#1\undefined}
}%
\providecommand \@ifnum [1]{%
 \ifnum #1\expandafter \@firstoftwo
 \else \expandafter \@secondoftwo
 \fi
}%
\providecommand \@ifx [1]{%
 \ifx #1\expandafter \@firstoftwo
 \else \expandafter \@secondoftwo
 \fi
}%
\providecommand \natexlab [1]{#1}%
\providecommand \enquote  [1]{``#1''}%
\providecommand \bibnamefont  [1]{#1}%
\providecommand \bibfnamefont [1]{#1}%
\providecommand \citenamefont [1]{#1}%
\providecommand \href@noop [0]{\@secondoftwo}%
\providecommand \href [0]{\begingroup \@sanitize@url \@href}%
\providecommand \@href[1]{\@@startlink{#1}\@@href}%
\providecommand \@@href[1]{\endgroup#1\@@endlink}%
\providecommand \@sanitize@url [0]{\catcode `\\12\catcode `\$12\catcode
  `\&12\catcode `\#12\catcode `\^12\catcode `\_12\catcode `\%12\relax}%
\providecommand \@@startlink[1]{}%
\providecommand \@@endlink[0]{}%
\providecommand \url  [0]{\begingroup\@sanitize@url \@url }%
\providecommand \@url [1]{\endgroup\@href {#1}{\urlprefix }}%
\providecommand \urlprefix  [0]{URL }%
\providecommand \Eprint [0]{\href }%
\@ifxundefined \urlstyle {%
  \providecommand \doi  [0]{\begingroup \@sanitize@url \@doi}%
  \providecommand \@doi [1]{\endgroup \@@startlink {\doibase
  #1}doi:\discretionary {}{}{}#1\@@endlink }%
}{%
  \providecommand \doi  [0]{doi:\discretionary{}{}{}\begingroup
  \urlstyle{rm}\Url }%
}%
\providecommand \doibase [0]{http://dx.doi.org/}%
\providecommand \Doi [0]{\begingroup \@sanitize@url \@Doi }%
\providecommand \@Doi  [1]{\endgroup\@@startlink{\doibase#1}\@@Doi}%
\providecommand \@@Doi [1]{#1\@@endlink}%
\providecommand \selectlanguage [0]{\@gobble}%
\providecommand \bibinfo  [0]{\@secondoftwo}%
\providecommand \bibfield  [0]{\@secondoftwo}%
\providecommand \translation [1]{[#1]}%
\providecommand \BibitemOpen [0]{}%
\providecommand \bibitemStop [0]{}%
\providecommand \bibitemNoStop [0]{.\EOS\space}%
\providecommand \EOS [0]{\spacefactor3000\relax}%
\providecommand \BibitemShut  [1]{\csname bibitem#1\endcsname}%
\bibitem [{\citenamefont {Drozdov}\ \emph {et~al.}()\citenamefont {Drozdov},
  \citenamefont {Eremets},\ and\ \citenamefont {Troyan}}]{Drozdov:2014}%
  \BibitemOpen
  \bibfield  {author} {\bibinfo {author} {\bibfnamefont {A.~P.}\ \bibnamefont
  {Drozdov}}, \bibinfo {author} {\bibfnamefont {M.~I.}\ \bibnamefont
  {Eremets}}, \ and\ \bibinfo {author} {\bibfnamefont {I.~A.}\ \bibnamefont
  {Troyan}},\ }\href@noop {} {\enquote {\bibinfo {title} {Conventional
  superconductivity at 190 $\rm k$ at high pressures},}\ }\Eprint
  {http://arxiv.org/abs/arXiv:1412.0460} {arXiv:1412.0460} \BibitemShut
  {NoStop}%
\bibitem [{\citenamefont {Duan}\ \emph {et~al.}(2014)\citenamefont {Duan},
  \citenamefont {Liu}, \citenamefont {Tian}, \citenamefont {Li}, \citenamefont
  {Huang}, \citenamefont {Zhao}, \citenamefont {Yu}, \citenamefont {Liu},
  \citenamefont {Tian},\ and\ \citenamefont {Cui}}]{Duan:2014}%
  \BibitemOpen
  \bibfield  {author} {\bibinfo {author} {\bibfnamefont {D.}~\bibnamefont
  {Duan}}, \bibinfo {author} {\bibfnamefont {Y.}~\bibnamefont {Liu}}, \bibinfo
  {author} {\bibfnamefont {F.}~\bibnamefont {Tian}}, \bibinfo {author}
  {\bibfnamefont {D.}~\bibnamefont {Li}}, \bibinfo {author} {\bibfnamefont
  {X.}~\bibnamefont {Huang}}, \bibinfo {author} {\bibfnamefont
  {Z.}~\bibnamefont {Zhao}}, \bibinfo {author} {\bibfnamefont {H.}~\bibnamefont
  {Yu}}, \bibinfo {author} {\bibfnamefont {B.}~\bibnamefont {Liu}}, \bibinfo
  {author} {\bibfnamefont {W.}~\bibnamefont {Tian}}, \ and\ \bibinfo {author}
  {\bibfnamefont {T.}~\bibnamefont {Cui}},\ }\Doi {10.1038/srep06968}
  {\bibfield  {journal} {\bibinfo  {journal} {Sci. Rep.},\ }\textbf {\bibinfo
  {volume} {4}},\ \bibinfo {pages} {6968} (\bibinfo {year} {2014})}\BibitemShut
  {NoStop}%
\bibitem [{\citenamefont {Errea}\ \emph {et~al.}(2015)\citenamefont {Errea},
  \citenamefont {Calandra}, \citenamefont {Pickard}, \citenamefont {Nelson},
  \citenamefont {Needs}, \citenamefont {Li}, \citenamefont {Liu}, \citenamefont
  {Zhang}, \citenamefont {Ma},\ and\ \citenamefont {Mauri}}]{Errea:2015}%
  \BibitemOpen
  \bibfield  {author} {\bibinfo {author} {\bibfnamefont {I.}~\bibnamefont
  {Errea}}, \bibinfo {author} {\bibfnamefont {M.}~\bibnamefont {Calandra}},
  \bibinfo {author} {\bibfnamefont {C.~J.}\ \bibnamefont {Pickard}}, \bibinfo
  {author} {\bibfnamefont {J.}~\bibnamefont {Nelson}}, \bibinfo {author}
  {\bibfnamefont {R.~J.}\ \bibnamefont {Needs}}, \bibinfo {author}
  {\bibfnamefont {Y.}~\bibnamefont {Li}}, \bibinfo {author} {\bibfnamefont
  {H.}~\bibnamefont {Liu}}, \bibinfo {author} {\bibfnamefont {Y.}~\bibnamefont
  {Zhang}}, \bibinfo {author} {\bibfnamefont {Y.}~\bibnamefont {Ma}}, \ and\
  \bibinfo {author} {\bibfnamefont {F.}~\bibnamefont {Mauri}},\ }\Doi
  {10.1103/PhysRevLett.114.157004} {\bibfield  {journal} {\bibinfo  {journal}
  {Phys. Rev. Lett.},\ }\textbf {\bibinfo {volume} {114}},\ \bibinfo {pages}
  {157004} (\bibinfo {year} {2015})}\BibitemShut {NoStop}%
\bibitem [{\citenamefont {Flores-Livas}\ \emph {et~al.}()\citenamefont
  {Flores-Livas}, \citenamefont {Sanna},\ and\ \citenamefont
  {Gross}}]{Flores-Livas:2015}%
  \BibitemOpen
  \bibfield  {author} {\bibinfo {author} {\bibfnamefont {J.~A.}\ \bibnamefont
  {Flores-Livas}}, \bibinfo {author} {\bibfnamefont {A.}~\bibnamefont {Sanna}},
  \ and\ \bibinfo {author} {\bibfnamefont {E.~K.~U.}\ \bibnamefont {Gross}},\
  }\href@noop {} {\enquote {\bibinfo {title} {High temperature
  superconductivity in sulfur and selenium hydrides at high pressure},}\
  }\Eprint {http://arxiv.org/abs/arXiv:1501.06336} {arXiv:1501.06336}
  \BibitemShut {NoStop}%
\bibitem [{\citenamefont {Bernstein}\ \emph {et~al.}(2015)\citenamefont
  {Bernstein}, \citenamefont {Hellberg}, \citenamefont {Johannes},
  \citenamefont {Mazin},\ and\ \citenamefont {Mehl}}]{Mazin:2015}%
  \BibitemOpen
  \bibfield  {author} {\bibinfo {author} {\bibfnamefont {N.}~\bibnamefont
  {Bernstein}}, \bibinfo {author} {\bibfnamefont {C.~S.}\ \bibnamefont
  {Hellberg}}, \bibinfo {author} {\bibfnamefont {M.~D.}\ \bibnamefont
  {Johannes}}, \bibinfo {author} {\bibfnamefont {I.~I.}\ \bibnamefont {Mazin}},
  \ and\ \bibinfo {author} {\bibfnamefont {M.~J.}\ \bibnamefont {Mehl}},\ }\Doi
  {10.1103/PhysRevB.91.060511} {\bibfield  {journal} {\bibinfo  {journal}
  {Phys. Rev. B},\ }\textbf {\bibinfo {volume} {91}},\ \bibinfo {pages}
  {060511} (\bibinfo {year} {2015})}\BibitemShut {NoStop}%
\bibitem [{\citenamefont {Papaconstantopoulos}\ \emph
  {et~al.}(2015)\citenamefont {Papaconstantopoulos}, \citenamefont {Klein},
  \citenamefont {Mehl},\ and\ \citenamefont {Pickett}}]{Pickett:2015}%
  \BibitemOpen
  \bibfield  {author} {\bibinfo {author} {\bibfnamefont {D.~A.}\ \bibnamefont
  {Papaconstantopoulos}}, \bibinfo {author} {\bibfnamefont {B.~M.}\
  \bibnamefont {Klein}}, \bibinfo {author} {\bibfnamefont {M.~J.}\ \bibnamefont
  {Mehl}}, \ and\ \bibinfo {author} {\bibfnamefont {W.~E.}\ \bibnamefont
  {Pickett}},\ }\href@noop {} {\bibfield  {journal} {\bibinfo  {journal} {Phys.
  Rev. B},\ }\textbf {\bibinfo {volume} {91}},\ \bibinfo {pages} {184511}
  (\bibinfo {year} {2015})}\BibitemShut {NoStop}%
\bibitem [{\citenamefont {Hirsch}\ and\ \citenamefont
  {Marsiglio}(2015)}]{Hirsch:2015}%
  \BibitemOpen
  \bibfield  {author} {\bibinfo {author} {\bibfnamefont {J.~E.}\ \bibnamefont
  {Hirsch}}\ and\ \bibinfo {author} {\bibfnamefont {F.}~\bibnamefont
  {Marsiglio}},\ }\href@noop {} {\bibfield  {journal} {\bibinfo  {journal}
  {Physica C},\ }\textbf {\bibinfo {volume} {511}},\ \bibinfo {pages} {45}
  (\bibinfo {year} {2015})}\BibitemShut {NoStop}%
\bibitem [{\citenamefont {Akashi}\ \emph {et~al.}()\citenamefont {Akashi},
  \citenamefont {Kawamura}, \citenamefont {Tsuneyuki}, \citenamefont {Nomura},\
  and\ \citenamefont {Arita}}]{Akashi:2015}%
  \BibitemOpen
  \bibfield  {author} {\bibinfo {author} {\bibfnamefont {R.}~\bibnamefont
  {Akashi}}, \bibinfo {author} {\bibfnamefont {M.}~\bibnamefont {Kawamura}},
  \bibinfo {author} {\bibfnamefont {S.}~\bibnamefont {Tsuneyuki}}, \bibinfo
  {author} {\bibfnamefont {Y.}~\bibnamefont {Nomura}}, \ and\ \bibinfo {author}
  {\bibfnamefont {R.}~\bibnamefont {Arita}},\ }\href@noop {} {\enquote
  {\bibinfo {title} {Fully non-emperical study on superconductivity in
  compressed sulfur hydrides},}\ }\Eprint
  {http://arxiv.org/abs/arXiv:1502.00936} {arXiv:1502.00936} \BibitemShut
  {NoStop}%
\bibitem [{\citenamefont {Carbotte}(1990)}]{Carbotte:1990}%
  \BibitemOpen
  \bibfield  {author} {\bibinfo {author} {\bibfnamefont {J.~P.}\ \bibnamefont
  {Carbotte}},\ }\Doi {10.1103/RevModPhys.62.1027} {\bibfield  {journal}
  {\bibinfo  {journal} {Rev. Mod. Phys.},\ }\textbf {\bibinfo {volume} {62}},\
  \bibinfo {pages} {1027} (\bibinfo {year} {1990})}\BibitemShut {NoStop}%
\bibitem [{sup()}]{supp}%
  \BibitemOpen
  \href@noop {} {}\bibinfo {note} {See Supplemental Material at [URL] for a
  comparison with other $\alpha^2F(\omega)$ spectra for H$_3$S and to see the
  optical conductivity curves at temperatures closer to $T_c$.}\BibitemShut
  {Stop}%
\bibitem [{\citenamefont {Allen}\ and\ \citenamefont
  {Dynes}(1975)}]{Allen:1975}%
  \BibitemOpen
  \bibfield  {author} {\bibinfo {author} {\bibfnamefont {P.~B.}\ \bibnamefont
  {Allen}}\ and\ \bibinfo {author} {\bibfnamefont {R.~C.}\ \bibnamefont
  {Dynes}},\ }\Doi {10.1103/PhysRevB.12.905} {\bibfield  {journal} {\bibinfo
  {journal} {Phys. Rev. B},\ }\textbf {\bibinfo {volume} {12}},\ \bibinfo
  {pages} {905} (\bibinfo {year} {1975})}\BibitemShut {NoStop}%
\bibitem [{\citenamefont {McMillan}(1968)}]{McMillan:1968}%
  \BibitemOpen
  \bibfield  {author} {\bibinfo {author} {\bibfnamefont {W.~L.}\ \bibnamefont
  {McMillan}},\ }\Doi {10.1103/PhysRev.167.331} {\bibfield  {journal} {\bibinfo
   {journal} {Phys. Rev.},\ }\textbf {\bibinfo {volume} {167}},\ \bibinfo
  {pages} {331} (\bibinfo {year} {1968})}\BibitemShut {NoStop}%
\bibitem [{\citenamefont {Marsiglio}\ \emph {et~al.}(1998)\citenamefont
  {Marsiglio}, \citenamefont {Startseva},\ and\ \citenamefont
  {Carbotte}}]{Marsiglio:1998}%
  \BibitemOpen
  \bibfield  {author} {\bibinfo {author} {\bibfnamefont {F.}~\bibnamefont
  {Marsiglio}}, \bibinfo {author} {\bibfnamefont {T.}~\bibnamefont
  {Startseva}}, \ and\ \bibinfo {author} {\bibfnamefont {J.~P.}\ \bibnamefont
  {Carbotte}},\ }\href@noop {} {\bibfield  {journal} {\bibinfo  {journal}
  {Phys. Lett. A},\ }\textbf {\bibinfo {volume} {245}},\ \bibinfo {pages} {172}
  (\bibinfo {year} {1998})}\BibitemShut {NoStop}%
\bibitem [{\citenamefont {Mitrovic}\ and\ \citenamefont
  {Carbotte}(1981)}]{Mitrovic:1981}%
  \BibitemOpen
  \bibfield  {author} {\bibinfo {author} {\bibfnamefont {B.}~\bibnamefont
  {Mitrovic}}\ and\ \bibinfo {author} {\bibfnamefont {J.~P.}\ \bibnamefont
  {Carbotte}},\ }\href@noop {} {\bibfield  {journal} {\bibinfo  {journal}
  {Solid State Commun.},\ }\textbf {\bibinfo {volume} {40}},\ \bibinfo {pages}
  {249} (\bibinfo {year} {1981})}\BibitemShut {NoStop}%
\bibitem [{\citenamefont {Carbotte}(1987)}]{Carbotte:1987}%
  \BibitemOpen
  \bibfield  {author} {\bibinfo {author} {\bibfnamefont {J.~P.}\ \bibnamefont
  {Carbotte}},\ }\href@noop {} {\bibfield  {journal} {\bibinfo  {journal} {Sci.
  Prog.},\ }\textbf {\bibinfo {volume} {71}},\ \bibinfo {pages} {329} (\bibinfo
  {year} {1987})}\BibitemShut {NoStop}%
\bibitem [{\citenamefont {Leavens}(1975)}]{Leavens:1975}%
  \BibitemOpen
  \bibfield  {author} {\bibinfo {author} {\bibfnamefont {C.~R.}\ \bibnamefont
  {Leavens}},\ }\href@noop {} {\bibfield  {journal} {\bibinfo  {journal} {Solid
  State Commun.},\ }\textbf {\bibinfo {volume} {17}},\ \bibinfo {pages} {1499}
  (\bibinfo {year} {1975})}\BibitemShut {NoStop}%
\bibitem [{\citenamefont {Leavens}\ and\ \citenamefont
  {Carbotte}(1974)}]{Leavens:1974}%
  \BibitemOpen
  \bibfield  {author} {\bibinfo {author} {\bibfnamefont {C.~R.}\ \bibnamefont
  {Leavens}}\ and\ \bibinfo {author} {\bibfnamefont {J.~P.}\ \bibnamefont
  {Carbotte}},\ }\href@noop {} {\bibfield  {journal} {\bibinfo  {journal} {J.
  Low Temp. Phys.},\ }\textbf {\bibinfo {volume} {14}},\ \bibinfo {pages} {195}
  (\bibinfo {year} {1974})}\BibitemShut {NoStop}%
\bibitem [{\citenamefont {Mitrovi\ifmmode~\acute{c}\else \'{c}\fi{}}\ \emph
  {et~al.}(1984)\citenamefont {Mitrovi\ifmmode~\acute{c}\else \'{c}\fi{}},
  \citenamefont {Zarate},\ and\ \citenamefont {Carbotte}}]{Mitrovic:1984}%
  \BibitemOpen
  \bibfield  {author} {\bibinfo {author} {\bibfnamefont {B.}~\bibnamefont
  {Mitrovi\ifmmode~\acute{c}\else \'{c}\fi{}}}, \bibinfo {author}
  {\bibfnamefont {H.~G.}\ \bibnamefont {Zarate}}, \ and\ \bibinfo {author}
  {\bibfnamefont {J.~P.}\ \bibnamefont {Carbotte}},\ }\Doi
  {10.1103/PhysRevB.29.184} {\bibfield  {journal} {\bibinfo  {journal} {Phys.
  Rev. B},\ }\textbf {\bibinfo {volume} {29}},\ \bibinfo {pages} {184}
  (\bibinfo {year} {1984})}\BibitemShut {NoStop}%
\bibitem [{\citenamefont {Geilikman}\ and\ \citenamefont
  {Kresin}(1966)}]{Geilikman:1966}%
  \BibitemOpen
  \bibfield  {author} {\bibinfo {author} {\bibfnamefont {B.~T.}\ \bibnamefont
  {Geilikman}}\ and\ \bibinfo {author} {\bibfnamefont {V.~Z.}\ \bibnamefont
  {Kresin}},\ }\href@noop {} {\bibfield  {journal} {\bibinfo  {journal} {Sov.
  Phys.-Solid State},\ }\textbf {\bibinfo {volume} {7}},\ \bibinfo {pages}
  {2659} (\bibinfo {year} {1966})}\BibitemShut {NoStop}%
\bibitem [{\citenamefont {Marsiglio}\ and\ \citenamefont
  {Carbotte}(1986)}]{Marsiglio:1986}%
  \BibitemOpen
  \bibfield  {author} {\bibinfo {author} {\bibfnamefont {F.}~\bibnamefont
  {Marsiglio}}\ and\ \bibinfo {author} {\bibfnamefont {J.~P.}\ \bibnamefont
  {Carbotte}},\ }\Doi {10.1103/PhysRevB.33.6141} {\bibfield  {journal}
  {\bibinfo  {journal} {Phys. Rev. B},\ }\textbf {\bibinfo {volume} {33}},\
  \bibinfo {pages} {6141} (\bibinfo {year} {1986})}\BibitemShut {NoStop}%
\bibitem [{\citenamefont {Akis}\ \emph {et~al.}(1991)\citenamefont {Akis},
  \citenamefont {Carbotte},\ and\ \citenamefont {Timusk}}]{Akis:1991}%
  \BibitemOpen
  \bibfield  {author} {\bibinfo {author} {\bibfnamefont {R.}~\bibnamefont
  {Akis}}, \bibinfo {author} {\bibfnamefont {J.~P.}\ \bibnamefont {Carbotte}},
  \ and\ \bibinfo {author} {\bibfnamefont {T.}~\bibnamefont {Timusk}},\ }\Doi
  {10.1103/PhysRevB.43.12804} {\bibfield  {journal} {\bibinfo  {journal} {Phys.
  Rev. B},\ }\textbf {\bibinfo {volume} {43}},\ \bibinfo {pages} {12804}
  (\bibinfo {year} {1991})}\BibitemShut {NoStop}%
\bibitem [{\citenamefont {Carbotte}\ \emph {et~al.}(1999)\citenamefont
  {Carbotte}, \citenamefont {Schachinger},\ and\ \citenamefont
  {Basov}}]{Carbotte:1999}%
  \BibitemOpen
  \bibfield  {author} {\bibinfo {author} {\bibfnamefont {J.~P.}\ \bibnamefont
  {Carbotte}}, \bibinfo {author} {\bibfnamefont {E.}~\bibnamefont
  {Schachinger}}, \ and\ \bibinfo {author} {\bibfnamefont {D.~N.}\ \bibnamefont
  {Basov}},\ }\href@noop {} {\bibfield  {journal} {\bibinfo  {journal} {Nature
  (London)},\ }\textbf {\bibinfo {volume} {401}},\ \bibinfo {pages} {354}
  (\bibinfo {year} {1999})}\BibitemShut {NoStop}%
\bibitem [{\citenamefont {Carbotte}\ \emph {et~al.}(2011)\citenamefont
  {Carbotte}, \citenamefont {Timusk},\ and\ \citenamefont
  {Hwang}}]{Carbotte:2011}%
  \BibitemOpen
  \bibfield  {author} {\bibinfo {author} {\bibfnamefont {J.~P.}\ \bibnamefont
  {Carbotte}}, \bibinfo {author} {\bibfnamefont {T.}~\bibnamefont {Timusk}}, \
  and\ \bibinfo {author} {\bibfnamefont {J.}~\bibnamefont {Hwang}},\
  }\href@noop {} {\bibfield  {journal} {\bibinfo  {journal} {Rep. Prog.
  Phys.},\ }\textbf {\bibinfo {volume} {74}},\ \bibinfo {pages} {066501}
  (\bibinfo {year} {2011})}\BibitemShut {NoStop}%
\bibitem [{\citenamefont {Mori}\ \emph {et~al.}(2008)\citenamefont {Mori},
  \citenamefont {Nicol}, \citenamefont {Shiizuka}, \citenamefont {Kuniyasu},
  \citenamefont {Nojima}, \citenamefont {Toyota},\ and\ \citenamefont
  {Carbotte}}]{Mori:2008}%
  \BibitemOpen
  \bibfield  {author} {\bibinfo {author} {\bibfnamefont {T.}~\bibnamefont
  {Mori}}, \bibinfo {author} {\bibfnamefont {E.~J.}\ \bibnamefont {Nicol}},
  \bibinfo {author} {\bibfnamefont {S.}~\bibnamefont {Shiizuka}}, \bibinfo
  {author} {\bibfnamefont {K.}~\bibnamefont {Kuniyasu}}, \bibinfo {author}
  {\bibfnamefont {T.}~\bibnamefont {Nojima}}, \bibinfo {author} {\bibfnamefont
  {N.}~\bibnamefont {Toyota}}, \ and\ \bibinfo {author} {\bibfnamefont {J.~P.}\
  \bibnamefont {Carbotte}},\ }\Doi {10.1103/PhysRevB.77.174515} {\bibfield
  {journal} {\bibinfo  {journal} {Phys. Rev. B},\ }\textbf {\bibinfo {volume}
  {77}},\ \bibinfo {pages} {174515} (\bibinfo {year} {2008})}\BibitemShut
  {NoStop}%
\end{thebibliography}%


\onecolumngrid

\pagebreak

\begin{center}
\textbf{\large Supplementary Information for\\
``Comparison of pressurized sulfur hydride with conventional superconductors''}
\end{center}

\begin{center}{E. J. Nicol$^{1}$ and J. P. Carbotte$^{2,3}$\\
{\it $^1$Department of Physics, University of Guelph,
Guelph, Ontario N1G 2W1, Canada\\ 
$^2$Department of Physics and Astronomy, McMaster
University, Hamilton, Ontario L8S 4M1, Canada\\
$^3$The Canadian Institute for Advanced Research, Toronto, ON M5G 1Z8, Canada}}
\end{center}

\twocolumngrid

\renewcommand{\thefigure}{S\arabic{figure}}
\renewcommand{\thetable}{S\arabic{table}}
\setcounter{figure}{0}


Shown in Fig.~\ref{fig1} are three of the $\alpha^2F(\omega)$ spectra 
available in the literature.
The
upper spectrum is a harmonic spectrum from Flores-Livas {\it et al.}[Ref.~4]
and the lower two are from Errea {\it et al.}[Ref.~3]. 
There is also a harmonic spectrum from Duan {\it et al.}[Ref.~2] which is not
shown here. These
spectra were digitized from the original papers and placed on a
2 meV grid which give characteristics that are in qualitative agreement
with the original works if not always in perfect quantitative agreement.
Such details do not impact our results and conclusions presented here.
It is clear that all of the spectra are qualitatively similar.
 \begin{figure}[ht]
\includegraphics[width=0.9\linewidth]{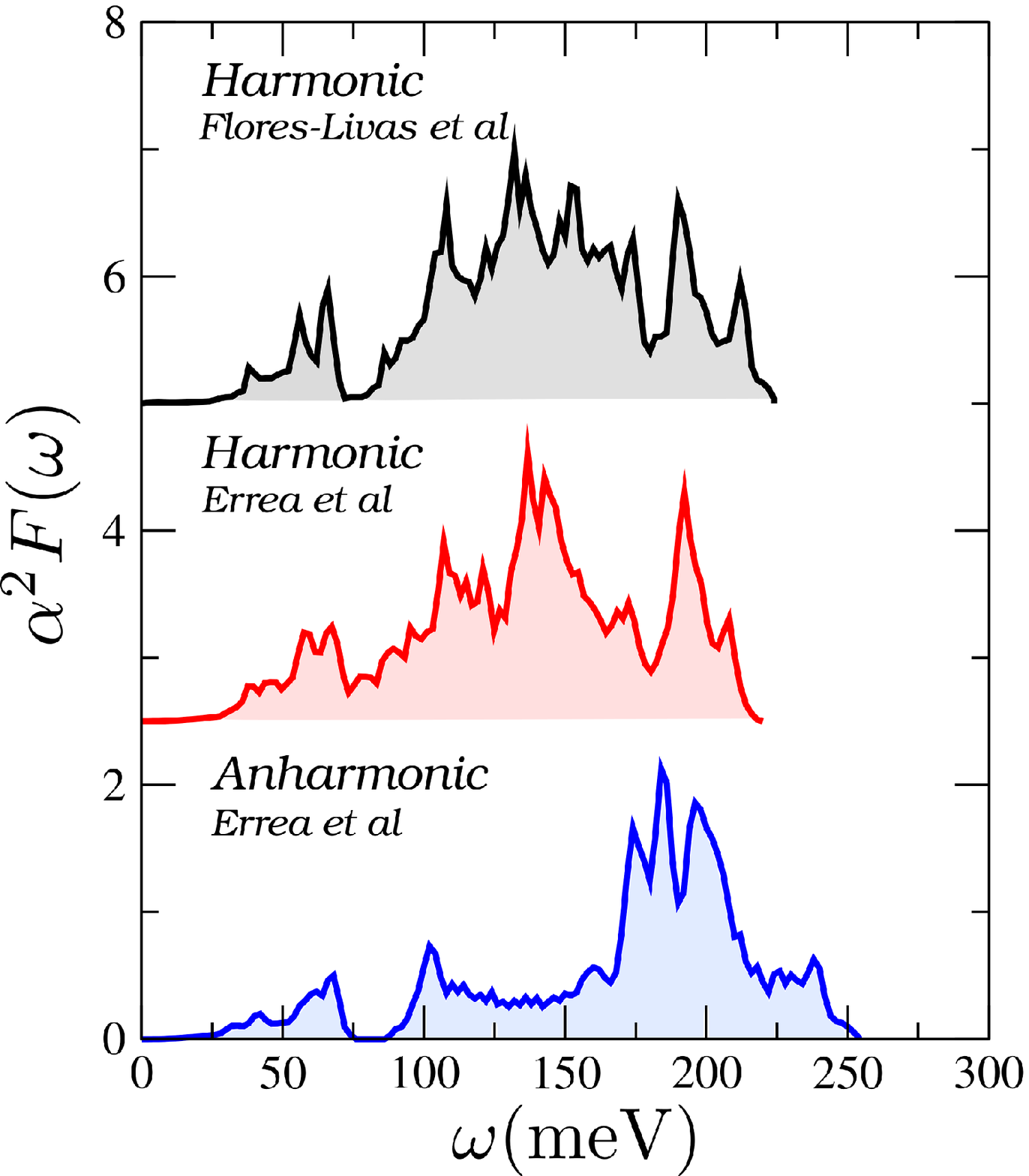}
\caption{(Color online) 
Three of the $\alpha^2F(\omega)$ spectra available in the
literature. The spectra have been offset for clarity. 
}\label{fig1}
\end{figure}

For our calculations, we have used the standard numerical solutions of the
imaginary axis s-wave Eliashberg equations in order to solve for $T_c$ and
other thermodynamic quantities.  By an analytic continuation to 
the real axis, we evaluated the energy gap and the optical conductivity.
A description of these techniques and formulas can be found in the review
by Carbotte [Ref.~9]. The optical conductivity calculation is done following the
procedure outlined by Akis et al [Ref.~21].

In the following table we show the characteristic parameters of our digitized
spectra. 

\begin{table}
\caption{Table of parameters for various digitized spectra}
\begin{ruledtabular}
\begin{tabular}{cccc}
$\alpha^2F(\omega)$  & Errea & Errea & Flores-Livas\\
& anharmonic &harmonic &harmonic \\
\hline\\
$\lambda$ & 1.67 & 2.55 & 2.47\\
$\omega_{\rm ln}$ (meV) &  122.1 & 102.6 & 106.1\\
$\omega_{\rm max}$ (meV) &  254 & 218 & 224\\
$A$ (meV) & 118.5 & 146.6 & 148.7\\
$T_c$ (K) & 190   & 190 (249)\footnotemark[1] &190 (255)\\
$\mu^*(N=6)$ & 0.18 & 0.38 (0.16) & 0.40 (0.16)\\
$T_c/\omega_{\rm ln}$ & 0.134   & 0.16 (0.209) &0.154 (0.207)\\
$\omega_{\rm ln}/T_c$ & 7.46   & 6.27 (4.79) & 6.48 (4.83)\\
$T_c/A$ & 0.138   & 0.112 (0.146) &0.110 (0.148)\\
$\Delta_0$ & 36.5   & 39.9 & 39.1\\
$2\Delta_0/k_BT_c$ & 4.46   & 4.87 & 4.78\\
\end{tabular}
\end{ruledtabular}
\footnotetext[1]{Entries in brackets correspond to taking $\mu^*=0.16$.}
\end{table}

In Fig.~\ref{fig2} we show the conductivity spectra for each of the
$\alpha^2F(\omega)$ shown in Fig.~\ref{fig1}.
To examine the effects of temperature, we show the conductivity for
the anharmonic spectrum at several temperatures in Fig.~\ref{fig3}.

\begin{figure}[h]
\includegraphics[width=0.9\linewidth]{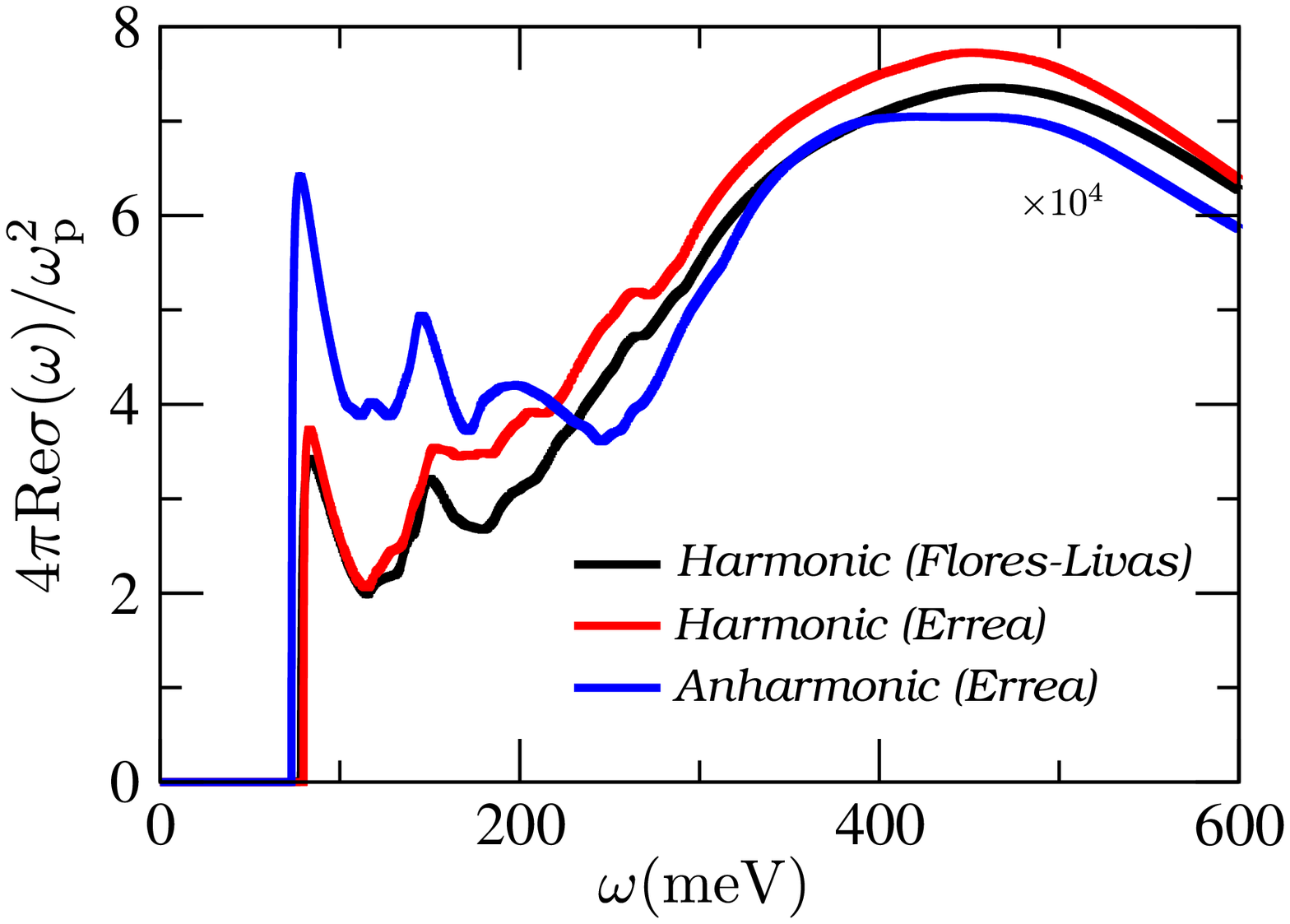}
\caption{(Color online)
A comparison of the real part of the conductivity in the superconducting
state for the three
different spectra shown in Fig.~\ref{fig1}. Here, $T=0.1T_c$.
}\label{fig2}
\end{figure}

\begin{figure}
\includegraphics[width=0.9\linewidth]{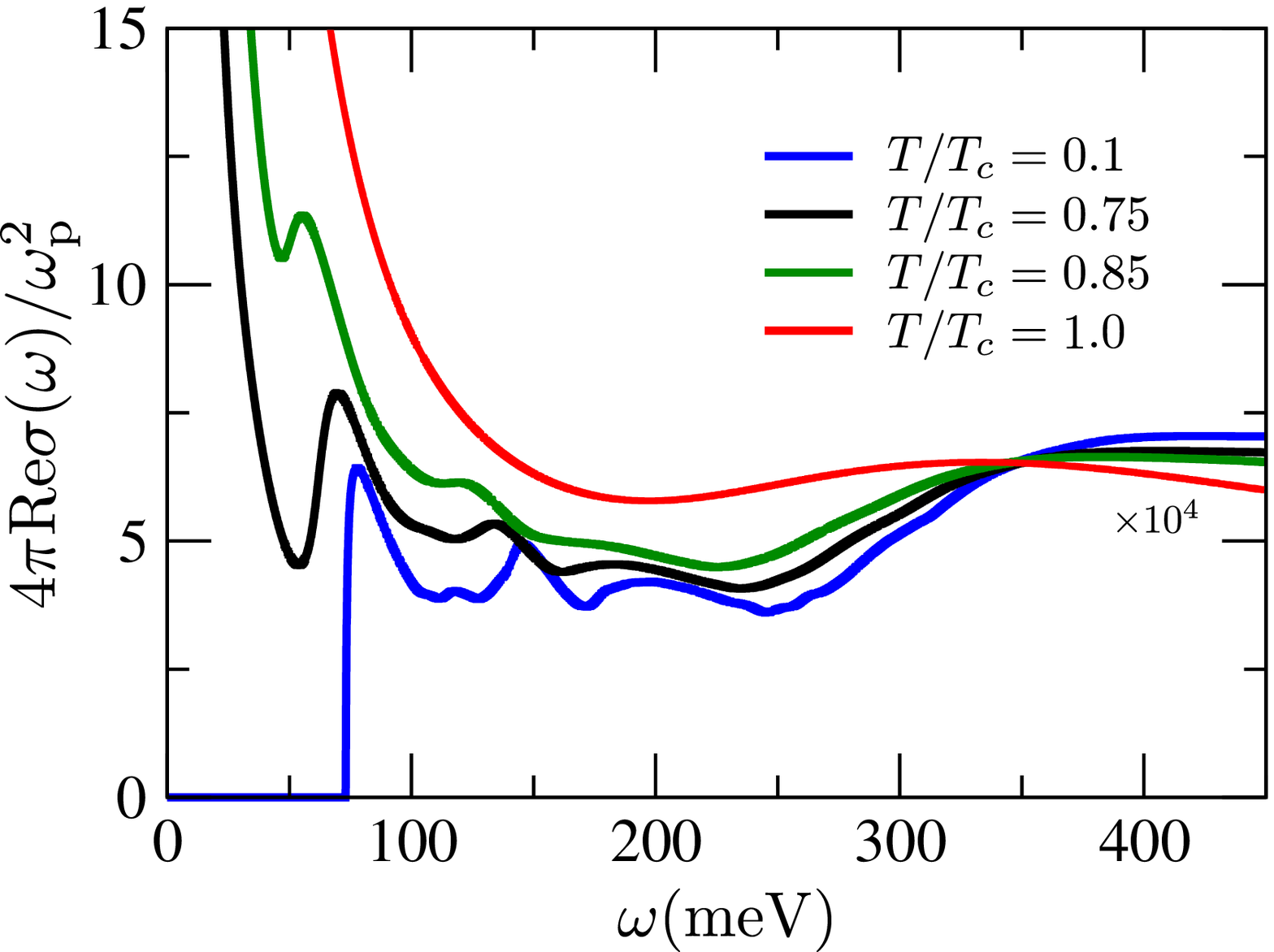}
\caption{(Color online) Variation of the real part of the
conductivity in the superconducting state for the anharmonic spectrum.
Curves are shown for different temperatures as indicated. The red curve is
for $T=T_c$ which is the normal state.
}\label{fig3}
\end{figure}

\end{document}